# INVESTIGATION OF DUST WAKE FIELD OSCILLATIONS

J. Kong*, T. Hyde$^\xi$, L. Matthews, M. Cook,
J. Schmoke, J. Carmona-Reyes

*CASPER, One Bear Place 97310,*
*Baylor University,* Waco, Texas 76798 USA

*Abstract*

Wakefield oscillations created by the ion wakefield existing below a dust particle within the plasma sheath generated above a powered lower electrode in a GEC rf reference cell carry information about the plasma sheath, the dust particle charge and the speed of the streaming ions. An experimental method to investigate such wakefield oscillations is discussed.

## I. INTRODUCTION

A wakefield is generally defined as a net positive space charge region residing downstream from a dust particle within a complex plasma. Wakefields are thought to be generated through an ion focusing effect created when positively charged ions stream from the plasma toward the lower electrode and are then reflected by suspended negatively charged dust particles [1, 2]. The resulting positive space-charge region gives rise to an attractive interaction between the negatively charged particles. One apparent effect of this attractive force is the alignment of the dust particles in the vertical direction, creating two- three- or even longer particle chains. Particles within such a chain are acted upon by the gravitational force, the sheath potential, the Yukawa potential, and the wake field potential. Therefore stable, vertical dust particle chains in complex plasmas provide an ideal platform for investigating the ion wakefield potential as well as the attractive/repulsive forces acting on the dust particles [3, 4].

A two-particle chain provides the simplest unit structure for such a particle chain system. In general, a more massive particle resides at the lower position within the chain with a less massive one directly above due to the gravitational force acting on the particles [3]. Mathematically a two-particle system has an exact set of solutions and recent computer simulations of the wakefield space charge distribution for this system predict an attractive force region between the two dust particles [5].

Experimentally, two-particle chains frequently appear spontaneously under typical laboratory dusty plasma conditions and are the simplest chain structures exhibiting the wakefield effect. The experiment discussed here is based on a resonance oscillation method. Employing a sine wave of frequency $\omega$ to modulate an externally applied DC bias to a powered lower electrode, an oscillation is generated in the particle chain system. The resulting oscillations of the two particles in the chain are out of phase, creating a relative velocity between them. The resulting motion provides information about the interaction forces.

## II. WAKE FIELD OSCILLATIONS

The wakefield potential downstream from a stationary dust particle can be theoretically described as [1]:

$$\phi_w(r=0,y) \approx \frac{Q_D}{4\pi\varepsilon_0 y}\frac{2\cos(y/L_s)}{1-M^{-2}} \quad (1)$$

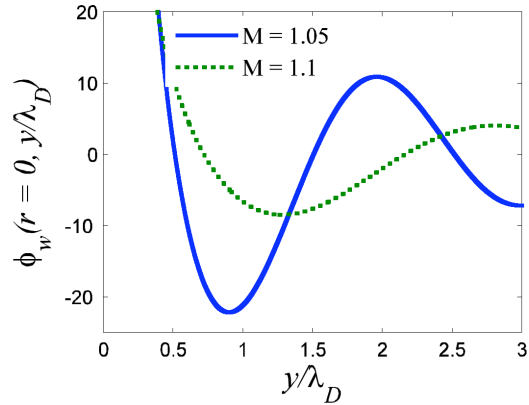

**Figure 1.** Wake potential at various Mach numbers.

In the above, $L_s = \lambda_D\sqrt{M^2-1}$ is the effective length, $Q_D$ is the dust charge, $M$ is the Mach number, $r$ is the cylindrical coordinate, and the origin of $y$ is assumed to be at the current position of the dust particle. The total potential acting on the dust particle in addition to the sheath potential is the sum of the wakefield and Yukawa potentials, $\phi = \phi_w + \phi_Y$. Fig. 1 shows $\phi$ to have a minimum at $y/\lambda_D = x_{\min}$, where this minimum is a function of $M$, $x_{\min} = x_{\min}(M)$. In order to examine the small amplitude vibrations of the dust particles within the potential well, $\phi$ is expanded as a Taylor series around the point $\dfrac{y}{\lambda_D} = x_{\min}$. This expansion yields,

$$\phi(r=0,y) = A_0 + A_1\left(\dfrac{y}{\lambda_D} - x_{\min}\right) + A_2\left(\dfrac{y}{\lambda_D} - x_{\min}\right)^2 + ... \quad (2)$$

---

* email: J_Kong@baylor.edu
$^\xi$ email: Truell_Hyde@baylor.edu

For small amplitude oscillations, terms past $O\left((y/\lambda_D - x_{\min})^3\right)$ can be ignored. The resulting electrical field $E = -\dfrac{\partial \phi}{\partial y}$ is,

$$E(r=0,y) = -\left(2A_2 \dfrac{y}{\lambda_D^2}\right) + C_0 \quad (3)$$

where $C_0$ is a constant.

The interaction force between the two dust particles is

$$F = Q_2 E = -Q_2 \dfrac{2A_2}{\lambda_D^2} y + C_1 \quad (4)$$

Where $C_1$ is a new constant. In the case of a second particle of mass $m_2$ and charge $Q_2$ located in this field, a small displacement from the equilibrium position will cause particle oscillations with a resonant frequency of

$$\omega_w^2 = \dfrac{2Q_2 A_2}{m_2 \lambda_D^2} = \dfrac{Q_2}{m_2} \dfrac{Q_1}{\varepsilon_0 \pi \lambda_D^3} M_f \quad (5)$$

where the above ignores a minor contribution from the Yukawa potential. As seen, the resonance frequency is also a function of the Mach number; therefore, if the value of $\omega_w$ can be obtained experimentally, the Mach number of the system at the dust particle's position can be derived.

For a typical dusty plasma experiment, the camera resolution is approximately $10\,\mu m$ and the Debye length, $\lambda_D$, is around $200\,\mu m$. Therefore, oscillations of the type discussed should be easily detectable if they occur at a frequency within the response range of the particle.

To evaluate the magnitude of $\omega_w$, equation (5) must be re-cast in the form,

$$\omega_w^2 = \omega_{02}^2 \dfrac{Z_1 M_f}{n_i \pi \lambda_D^3} \quad (6)$$

where $Q_1 = Z_1 e$, $n_i$ is the ion density and $\omega_{02}$ is the lower particle's resonance frequency. Again assuming typical dusty plasma experimental parameters, $Z_1 = 6\times 10^3$, $n_i = \varepsilon_0 K_B T_i / e^2 \lambda_D^2$, $K_B T_i = 0.03\,eV$, $\lambda_D = 200-300\,\mu m$ and $Z_1/n_i \pi \lambda_D^3 \approx 5$, Equation (6) yields $\omega_w \approx 2.2 \omega_{02} \sqrt{M_f}$. Fig. 2 shows $\sqrt{M_f}$ as a function of $M$ for non-negative values of $M$. ($M_f$ is negative for $M<1$, thus, there can be no oscillations in this regime. Therefore, only when the Mach number is greater than one will it be possible to induce a wakefield oscillation.) Experimentally, only the lower particle in a two-particle chain is acted upon by the wake field potential; therefore, comparing the oscillation spectra of the two particles provides the data necessary to extract any induced wake field oscillations.

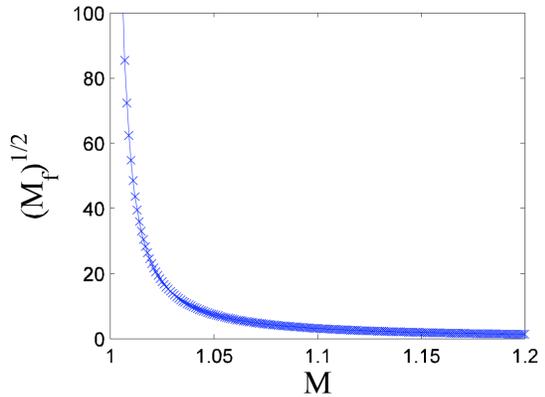

**Figure 2.** $\sqrt{M_f}$ as a function of $M$. $M_f$ increases rapidly as $M \to 1$.

### III. EXPERIMENTAL SETUP

The experiment described herein was conducted employing the CASPER GEC rf reference cell [6]. A radio-frequency, capacitively coupled discharge was formed between two parallel-plate electrodes, 8 cm in diameter and separated by 3 cm, with the bottom electrode air-cooled. The lower electrode is powered by a radio-frequency signal generator, while the upper electrode is grounded as is the chamber. The signal generator is coupled to the electrode through an impedance matching network and a variable capacitor attenuator network. The plasma discharge apparatus is described in greater detail in [7].

To expedite data collection, multiple frequencies with identical amplitude were employed to drive the lower electrode, where a Fourier transformation of the particle response was then used to obtain the final frequency spectrum. Input voltages were calculated as a function of time

$$V_{in} = A \sum_{f_i = f_1}^{f_N} \sin(2\pi f_i t) \qquad (7)$$

where in the above, A is a constant. Fig. 3 shows $V_{in}$ as a function of time along with its corresponding fast Fourier transformation. In the figure, $f_i$ runs from $10 Hz$ to $20 Hz$ with a step size of $1 Hz$. These calculated $V_{in}$ values, along with a time step size of $\Delta t = 1/1200\, s$, were fed to the lower electrode through an Agilent 33120A function generator in order to modulate the DC voltage. Fig. 4 shows the vertical response for two particles driven by the above input voltage, $V_{in}$, along with their corresponding fast Fourier transformation. Particle position was recorded using a CCD camera at 120 frames per second for an argon gas plasma held at 100mTorr under a rf power of 5W and frequency of 13.56 MHz. The resulting data allows the resonance frequency $\omega_0$, and damping coefficient $\beta$ to be derived by fitting the data theoretically.

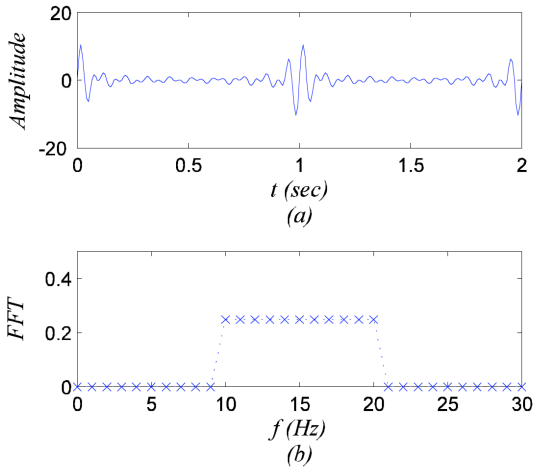

**Figure 3.** Input driving voltage waveform and corresponding Fourier transformation. (a). Driving waveform as a function of time. (b). Fast Fourier transformation.

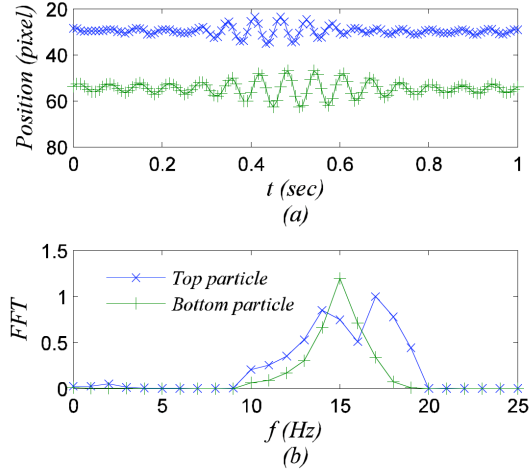

**Figure 4.** Particle response under the driving waveform shown in Fig. 3 and described in the text. (a). Particle response. (b). Fast Fourier transformation of (a).

In order to verify these results, a second experiment was conducted to measure particle resonance frequencies by varying the DC bias on the lower electrode as a mechanism for raising the particles to a height $\Delta h$ above their natural bias position, and then removing the external DC bias. Upon release, particles oscillate with attenuated amplitudes until returning to their natural bias positions. Fig. 5 shows this process for a two-particle pair along with the corresponding Fourier transformations. Experimental conditions are identical to the previous case discussed other than the initial DC bias being held negatively at a value 10V higher than the natural bias.

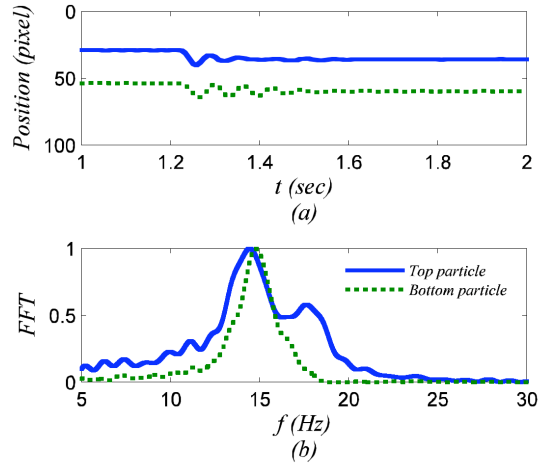

**Figure 5.** Attenuated oscillations of a two-particle chain and corresponding Fourier transformations.

In the above, particle charges are derived from measured resonant frequencies. The ion density $n_i$ is related to the Debye length $\lambda_D$ by,

$$\frac{n_i e}{\varepsilon_0} = \frac{K_B T_i}{e \lambda_D^2} \tag{8}$$

where $K_B$ is the Boltzmann constant, and $T_i$ is the ion temperature. $\lambda_D$ is derived from an experimental method following one developed in [8].

## IV. RESULTS AND DISCUSSIONS

Figs. 4(b) and Fig. 5(b) show the spectrum for the upper particle to have a very strong peak at the lower particle's resonance frequency, while the lower particle's spectrum shows only a weak peak at the top particle's resonance frequency. The explanation for this phenomenon can be attributed to the interaction forces, where both Yukawa and wakefield are considered. As can be seen in the figures, the bottom particle's vibration amplitude is larger, and the vibration duration time is longer. Both of these indicate that the bottom particle is being acted upon by a smaller frictional coefficient. Additionally, the phase difference between the oscillations of the two particles create a non-stable relative separation distance as can be seen in both particles' Fourier transformation spectrum. Assuming the interaction strength felt by both particles to be the same, the relative amplitude of the Fourier transformation spectrum will be dependent upon the amplitudes of the vibration. This is manifested in the top particle's Fourier transformation spectrum showing a strong interaction peak, while the bottom spectrum shows only a relatively weak one.

Since the wakefield is formed downstream to a standing particle, only the bottom particle should be acted upon by the potential. Oscillations within the wakefield potential well can be created by small displacements from equilibrium; since the bottom of the wakefield potential well (fig. 1), will move along with the top particle, only when the two particles undergo relative motion should there be an oscillation in the wakefield potential described in equation (5). A plan for observing such possible oscillations will be described in a subsequent paper.

## V. SUMMARY

The wakefield potential generated by a dust particle residing in the sheath of a plasma has been previously shown to create an attractive force on the particle underneath. Theoretical wakefield potential distributions imply the existence of a potential well within this potential, providing a possible oscillation field for dust particles trapped in the well. The oscillation frequency for such a well was derived in this paper, while the conditions necessary for the experimental observation of any resulting oscillations were discussed. The wakefield oscillation frequency was shown to be related to the Mach number of the system, providing an experimental method for determination. The wakefield oscillation frequency was also shown to depend on the charge and mass of the dust particles and this was used to provide a possible experimental method for measuring the mass and charge difference between them. An experiment was conducted employing an external DC bias to create a perturbation between the particles in a two-particle chain and the data analyzed based on a Fourier transformation method. Although this experiment did not show wakefield oscillations, it did introduce a new technique for investigating the interaction forces between the dust particles.